
\magnification\magstep0
\hsize = 12.0 cm
\vsize = 18.0 cm
\hoffset=0.70 true cm
\voffset=1.10 true cm
%
%
\nopagenumbers
\tenrm

\noindent{\bf Coadsorption of Copper and Sulfate on Au(111) Electrodes:}
\vskip 0.0 true cm
\noindent{\bf Monte Carlo Simulation of a Lattice-Gas
Model}\footnote*{To appear in {\it Computer Simulation Studies
in Condensed-Matter Physics VIII},
edited by D.~P. Landau, K.~K. Mon, and H.~B. Sch{\"u}ttler
(Springer, Berlin, in press).}

\vskip  0.15 true cm
\noindent{Jun Zhang$^1$, Per Arne Rikvold$^{1,2}$, Yung-Eun Sung$^3$
and Andrzej Wieckowski$^3$}
\vskip  0.15 true cm
\noindent{$^1$Supercomputer Computations Research Institute,}
\vskip -0.05 true cm
\hskip -0.55 true cm
Florida State University, Tallahassee, FL ~32306-4052, ~USA

\noindent{$^2$Department of Physics and Center for Materials Research and
Technology,}
\vskip -0.05 true cm
\hskip -0.55 true cm
Florida State University, Tallahassee, FL 32306-3016, ~USA

\noindent{$^3$Department of Chemistry, University of Illinois,
Urbana, IL ~61801, ~USA}

\vskip 0.5 true cm
\noindent{\bf Abstract.}
We report ground-state calculations and Monte Carlo simulations for
a lattice-gas model of the underpotential deposition of copper on
Au(111) in sulfate-containing electrolytes.
In a potential range of approximately 100$\sim$150 mV,
this system exhibits a
$(\sqrt3\!\times\!\sqrt3)$ mixed phase with 2/3 monolayer (ML) copper and 1/3
ML sulfate.
Our simulation results agree well with experimental results and with
other theoretical work.

\vskip 0.2 true cm
\noindent{\bf 1. Introduction}
\vskip 0.1 true cm
\noindent
Underpotential deposition (UPD) is a process whereby a monolayer (ML) or
less of one metal is electrochemically adsorbed onto another in a range of
electrode potentials more positive than those where bulk deposition would
occur [1]. The UPD of copper on Au(111) electrodes in
sulfate-containing electrolytes has been intensively studied,
both experimentally (see detailed discussion in Ref.~[2])
and theoretically [3,4].
In cyclic voltammetry (CV) experiments the potential is scanned
slowly while the current density is recorded.
The most striking feature of the CV currents observed with a Au(111)
electrode in sulfate-containing electrolyte
is the appearance of two peaks,
separated by about 100$\sim$150 mV, upon the addition of Cu$^{2+}$ ions
[5,6].
Typical CV profiles are shown in Fig.~1a [2], together with our simulation
results.
In the potential range between the peaks, the adsorbate layer is believed to
have a ($\sqrt3$$\times$$\sqrt3$) structure consisting of 2/3 ML
copper and 1/3~ML sulfate. This picture is based on a variety of both
{\it ex situ}\ and {\it in situ}\
experimental evidence, as discussed in Ref.~[2].

\vskip 0.2 true cm
\noindent{\bf 2. Model and theoretical methods}
\vskip 0.1 true cm
\noindent
Our lattice-gas model for UPD of copper on Au(111) in sulfate-containing
electrolyte is a refinement of the Huckaby-Blum model [3,4].
It is defined by a standard three-state lattice-gas Hamiltonian [7,8],
$$ \eqalign{
{\cal H}_{\rm LG} =
& \sum_l \bigl[ -\Phi_{\rm SS}^{(l)}
        \sum_{\langle ij \rangle}^{(l)} n_i^{\rm S} n_j^{\rm S}
       -\Phi_{\rm SC}^{(l)} \sum_{\langle ij \rangle}^{(l)}
        \left(n_i^{\rm S} n_j^{\rm C}
       + n_i^{\rm C} n_j^{\rm S} \right)
       -\Phi_{\rm CC}^{(l)}
        \sum_{\langle ij \rangle}^{(l)} n_i^{\rm C} n_j^{\rm C} \bigr] \cr
&  - \Phi_{\rm SS}^{(t)}\sum_\bigtriangleup n_in_jn_k
    - \bar{\mu}_{\rm S} \sum_i n_i^{\rm S}
    - \bar{\mu}_{\rm C} \sum_i n_i^{\rm C}  \; . \cr} \eqno (1)
$$
Here $n_i^{\rm X}\in\{0,1\}$
is the local occupation variable for species X
[X=S (sulfate) or~C (copper atom)], and the third adsorption state
(``empty'' or ``solvated'')
corresponds to $n_i^{\rm S}$=$n_i^{\rm C}$=0.
The sums $\sum_{\langle ij \rangle}^{(l)}$, $\sum_\bigtriangleup$,
and $\sum_i$ run over all
$l$th-neighbor bonds, over all equilateral next-nearest-neighbor
triangles [9], and over all adsorption sites,
respectively, $\Phi_{\rm XY}^{(l)}$ denotes the
effective XY interaction through an
$l$th-neighbor bond, and $\sum_l$ runs over the interaction ranges.
The change in electrochemical potential when one X
particle is removed from the bulk solution and adsorbed on the surface
is $-\bar{\mu}_{\rm X}$.
The sign convention is such that $\Phi_{\rm XY}^{(l)}$$>$0
denotes an effective attraction, and $\bar{\mu}_{\rm X}$$>$0
denotes a tendency for adsorption in the absence of lateral
interactions.
The bonds that correspond to finite lateral interaction energies are shown in
Fig.~2. For large separations the
interactions vanish,
and $\Phi_{\rm SS}^{(1)}$ is an infinite repulsion corresponding
to nearest-neighbor sulfate-sulfate exclusion
(``hard hexagons" [3,4,10]).
We emphasize that the $\Phi_{\rm XY}^{(l)}$ are {\it effective}
interactions through several channels, including electron-, phonon-,
and fluid-mediated mechanisms [7].

The electrochemical potentials in Eq.~(1) are (in the weak-solution
approximation and here given in molar units) related to the bulk
concentrations [X] and the electrode potential $E$ as
$$
\bar{\mu}_{\rm X} = \bar{\mu}_{\rm X}^0 + RT\ln {[{\rm X}]\over[{\rm X}]^0}
- z_{\rm X}FE \; . \qquad
\eqno (2) $$
Here $z_{\rm X}$ (X = S,C) are the effective electrovalences of
sulfate and copper, $R$ is the molar gas constant, $T$ is the
absolute temperature, and $F$ is the Faraday constant. The quantities
superscripted with a 0 are reference values which contain the local binding
energies to the surface. They are generally temperature dependent if the
effects of rotational and vibrational modes are considered.

The coverages of sulfate and copper
are defined as $\theta_{\rm X} = N^{-1}\sum n_i^{\rm X}$, where $N$ is the
total number of unit cells in the lattice.
However, it is experimentally observed that sulfate remains adsorbed
on top of the copper monolayer
in the negative-potential region, rather than becoming reduced and
entering the solution. A simple estimate of the sulfate coverage in this
second layer can be obtained as
$\theta_{\rm S}^{(2)} = \alpha\theta_{\rm C}(1/3-\theta_{\rm S})$,
which allows the difference between the first-layer coverage $\theta_{\rm S}$
and its saturation value of 1/3 to be transferred to the top of the copper
layer. The factor $\alpha$ is a phenomenological constant expected
to be between zero and one.
Since the transfer of sulfate between the gold and copper surfaces does not
involve an oxidation/reduction process, the total charge transport
per unit cell during the adsorption/desorption process
is $q = -e [ z_{\rm s} ( \theta_{\rm S}$+$\theta_{\rm S}^{(2)} )
+ z_{\rm C} \theta_{\rm C} ]$, where $e$ is the elementary charge.
In the absence of diffusion and double-layer effects and in the limit that
the potential scan rate ${\rm d}E/{\rm d}t\!
\rightarrow\! 0$ [1], the voltametric
current $i$ per unit cell of the surface is the time derivative of $q$.
Using differentiation by parts involving the relations between the $E$
and $\bar{\mu}_{\rm X}$, Eq.~(2), as well
as the Maxwell relation
$\partial \theta_{\rm S}/\partial \bar{\mu}_{\rm C} =
\partial \theta_{\rm C}/\partial \bar{\mu}_{\rm S}$,
we find the current density $i$ in terms of the lattice-gas response functions:
$$ \eqalign{
i = & eF\Bigl\{ z_{\rm S}^2(1-\alpha\theta_{\rm C})
{\partial\theta_{\rm S}\over\partial\bar{\mu}_{\rm S}}
\Bigm|_{\bar{\mu}_{\rm C}}
+z_{\rm C}(z_{\rm C}-2\alpha z_{\rm S}\theta_{\rm S}/3)
{\partial \theta_{\rm C}\over\partial \bar{\mu}_{\rm C}}
\Bigm|_{\bar{\mu}_{\rm S}}   \cr
  &    +
z_{\rm S}\Bigl(2z_{\rm C}+\alpha z_{\rm S}(1/3-\theta_{\rm S})
-\alpha z_{\rm C}\theta_{\rm C}\Bigr)
{\partial\theta_{\rm S}\over\partial \bar{\mu}_{\rm C}}
\Bigm|_{\bar{\mu}_{\rm S}}\Bigr\}
{{\rm d}E\over{\rm d}t}\,, \cr}
\eqno (3) $$
which reduces to its standard form for $\alpha$=0 [11].

Although the experimental studies
are carried out at room temperature, the zero-temperature
phase diagram serves as a quite accurate guide to the path
in the ($\bar{\mu}_{\rm S}$,$\bar{\mu}_{\rm C}$) plane that the
isotherms should follow.
At constant temperature and $p$H,
two factors influence the path: the adsorbate concentrations in the
electrolyte and the electrovalences.
As seen from Eq.~(2), $\bar{\mu}_{\rm S}$ and
$\bar{\mu}_{\rm C}$ depend linearly on $E$, with slopes
determined by $z_{\rm S}$ and $z_{\rm C}$, whose
values must be determined from experiments.
Here we use $z_{\rm C}$=+2, $z_{\rm S}$=$-$2.
Thirty-two ordered phases were found by applying the group-theoretical
arguments of Landau and Lifshitz [12], nine of which
(denoted by ($X$$\times$$Y$)$_{\theta _{\rm C}}^{\theta _{\rm S}}$ in Fig.~3)
are realized as
ground states for interactions in the region of experimental interest.
The ground-state energies
depend on $\bar{\mu}_{\rm S}$ and $\bar{\mu}_{\rm C}$ and the lateral
interactions.
The repulsive second-neighbor three-particle interaction $\Phi_{\rm SS}^{(t)}$
disfavors the
pure sulfate $(\sqrt{3}\!\times\!\sqrt{3})_0^{1/3}$ phase, which has not been
experimentally observed in this system. For a fixed set of
interactions, the zero-temperature phase boundaries are exactly determined
by pairwise equating the ground-state energies. In order to easily
explore the effects of changing the interactions, a program was
developed which numerically determines the zero-temperature phase diagram by
scanning $\bar{\mu}_{\rm S}$ and $\bar{\mu}_{\rm C}$ and determining the
phase of minimum energy [13].

To obtain adsorption isotherms and CV currents at room temperature, we
performed Monte Carlo (MC) simulations on a 30$\times$30 triangular lattice,
using a heat-bath algorithm [14]
with updates at randomly chosen sites. In order to avoid getting stuck in
metastable configurations (a problem which is exacerbated by the
nearest-neighbor sulfate-sulfate exclusion), we simultaneously updated clusters
consisting of two nearest-neighbor sites.
Each data point was obtained as an average over 2$\times 10^5$ Monte Carlo
Steps per spin (MCSS), sampling at intervals of 50 MCSS and discarding the
first 4000 MCSS.

\vskip 0.2 true cm
\noindent{\bf 3. Numerical results}
\vskip 0.1 true cm
\noindent
The zero-temperature phase diagram corresponding to the interactions used
in this work is shown in Fig.~3. For large negative $\bar{\mu}_{\rm S}$
only copper adsorption is possible,
and the phase diagram is that of the lattice-gas model corresponding to the
triangular-lattice antiferromagnet with next-nearest neighbor ferromagnetic
interactions [15]. Similarly, in the limit of large positive
$\bar{\mu}_{\rm S}$ and large negative $\bar{\mu}_{\rm C}$ the
zero-temperature phase is the
$(\sqrt{3}\!\times\!\sqrt{3})_0^{1/3}$ sulfate phase characteristic
of the hard-hexagon model [3,4,10].
The phase diagram for intermediate electrochemical
potentials is quite complicated. For $\bar{\mu}_{\rm S} \! < \! -22$~kJ/mol,
no sulfate adsorption occurs in the first adlayer,
while if $\bar{\mu}_{\rm C} \! < \! -18$ kJ/mol, no copper is adsorbed.
The $(\sqrt{3}\!\times\!\sqrt{7})_0^{1/5}$ phase corresponds to experimental
observations in copper-free systems [16,17]. It is enhanced by
the fourth-neighbor sulfate-sulfate attraction, $\Phi_{\rm SS}^{(4)}$, and
the $(\sqrt{3}\!\times\!\sqrt{3})_0^{1/3}$
phase is disfavored
by the second-neighbor repulsive trios, $\Phi_{\rm SS}^{(t)}$.
The $(\sqrt{3}\!\times\!\sqrt{3})_{2/3}^{1/3}$
mixed-phase region in the upper right-hand part of the diagram is
relatively large,
due to the nearest-neighbor attraction between copper and sulfate,
$\Phi_{\rm SC}^{(1)}$.
The isothermal path is chosen such that the distance between points A and B
(measured by the electrode potential) equals the peak separation in the
CV current shown in
Fig.~1a. The narrow strip of
$(\sqrt{3}\!\times\!\sqrt{7})_{4/5}^{1/5}$
phase lies closely above the $(\sqrt{3}\!\times\!\sqrt{3})_{2/3}^{1/3}$ phase
and is quite sensitive to the fourth-neighbor
attraction, $\Phi_{\rm SS}^{(4)}$, and the second-neighbor trio sulfate
repulsion,
$\Phi_{\rm SS}^{(t)}$.
The nearest-neighbor copper repulsion causes the appearance of
the $(\sqrt{3}\!\times\!\sqrt{3})_{1/3}^{1/3}$ phase.

The potential scan path corresponding to the CV current and
adsorption isotherms
shown in Fig.~1 is indicated by the dashed line in Fig.~3.
With the aid of the ground-state diagram, it is easy to analyse the simulation
results.
Starting from the positive potential end (lower right in Fig.~3),
where $\theta_{\rm C}\!\approx\!0$,
$\theta_{\rm S}$ drops from its saturated hard-hexagon value of 1/3 to
approximately 1/5 in the $(\sqrt{3}\!\times\!\sqrt{7})_0^{1/5}$ phase region.
This is in reasonable agreement
with the experimental results for the same
electrochemical system in the absence of copper, where the
$(\sqrt{3}\!\times\!\sqrt{7})_0^{1/5}$ structure has been
observed, but not that of
$(\sqrt{3}\!\times\!\sqrt{3})_0^{1/3}$ [16,17].
As $E$ is scanned in the negative direction, the
cations are attracted toward the electrode. Since there are
strong effective
attractions between the two types of adsorbed particles,
the adsorption of copper
induces the readsorption of sulfate: they both increase their
coverages in a narrow potential range near 175 mV to form the mixed
$(\sqrt{3}\!\times\!\sqrt{3})_{2/3}^{1/3}$ phase. This is exactly the scenario
proposed by Huckaby and Blum [3,4], and it is an example
of the enhanced-adsorption phenomenon discussed by
Rikvold and Deakin [7]. This phase remains stable in a potential
range indicated by the separation of the two CV peaks,
until the electrode potential is sufficiently
negative that the sulfate adsorption on the gold surface
is disfavored in comparison
with completion of the copper monolayer.
The replacement causes another sharp change in the
surface coverages, corresponding to the left CV peak.
However, there is experimental evidence that part
of the sulfate
desorbed from the gold surface is not reduced and dissolved, but rather
remains adsorbed in a
formally neutral submonolayer on top of the monolayer of copper, with a
coverage $\theta_{\rm S}^{(2)} \! \approx \! 0.2$ [6]. This corresponds
to $\alpha$=0.6 in Eq.~(3), which was used to obtain the simulated CV
current and surface-charge densities shown in Fig.~1.
Figure~1b shows a
comparison of the simulation results for the charge transfer and the
integral of the experimental CV current.

The agreement between the experimental and theoretical results is reasonable,
except for large positive $E$, where the model predicts less copper and more
sulfate on the surface than indicated by the experiments.
The disagreement between the
theoretical and experimental maximum currents may be due to defects on the
electrodes used in the experiments.

\vskip 0.2 true cm
\noindent{\bf Acknowledgements}
\vskip 0.1 true cm
\noindent
We have enjoyed enlightening conversations with L. Blum.
This work was supported by Florida State University
through the Supercomputer Computations Research Institute
(funded in part by
US Department of Energy Contract No.\ DE-FC05-85-ER25000) and the Center for
Materials Research and Technology, and by NSF grant No.\ DMR-9315969.
Work at the University of Illinois was supported by the Frederick Seitz
Materials Research Laboratory under US Department of Energy Contract No.\
DE-AC02-76-ER01198.

\vfill\eject
\centerline{\bf Figure Captions}
\vskip 1.5 true cm

\noindent {\bf Figure 1:}~
Experimental (dashed curves) and simulated (solid curves) results.
(a): CV currents for
a positive-going potential scan at 2 mV/s. (b): Integrated charge density.
Also shown are $z_{\rm C}\theta_{\rm C}$ (long dashed),
$z_{\rm S}\theta_{\rm S}$ (dotted) and
$z_{\rm S}(\theta_{\rm S}+\theta_{\rm S}^{(2)})$ (dotdash).
\vskip 1.5 true cm

\noindent {\bf Figure 2:}~
The relative positions of copper (filled circles) and sulfate (triangles)
corresponding to the effective interactions used in Eq.~(1). The
numbers are the corresponding values
of $\Phi_{\rm XY}^{(l)}$ used in this work, given in kJ/mol. The interactions
are invariant under symmetry operations on the lattice.
\vskip 1.5 true cm

\noindent {\bf Figure 3:}~
The zero-temperature phase diagram. Solid lines are phase boundaries and the
dotted line is the path along which the isotherms are calculated (positive E
towards the lower right). The phases are indicated as
($X$$\times$$Y$)$_{\theta _{\rm C}}^{\theta _{\rm S}}$.
\vskip 1.5 true cm

\vfill
\eject

\centerline{\bf References}
\vskip 0.1 true cm
\item{[1]}
A.J.\ Bard and L.R.\ Faulkner, {\it Electrochemical Methods}
(Wiley, New York, 1985).

\item{[2]}
J.\ Zhang, Y.-E.\ Sung, P.A.\ Rikvold, and A.\ Wieckowski,
to be submitted to Surf.\ Sci.

\item{[3]}
D.A.\ Huckaby and L.~Blum,
J.\ Electroanal.\ Chem.\ {\bf 315}, 255 (1991).

\item{[4]}
L.~Blum and D.A.\ Huckaby,
J.\ Electroanal.\ Chem.\ {\bf 375}, 69 (1994).

\item{[5]}
M.~Zei, G.~Qiao, G.~Lempfuhl, and D.M.\ Kolb,
Ber.\ Bunsenges.\ Phys.\ Chem.\ {\bf 91}, 3494 (1987).

\item{[6]}
Z.~Shi and J.~Lipkowski,
J.\ Electroanal.\ Chem.\ {\bf 364}, 289 (1993); {\bf 365}, 303 (1993).

\item{[7]}
P.A.\ Rikvold and Mark~R.\ Deakin,
Surf. Sci. {\bf 249}, 180 (1991).

\item{[8]}
P.A.\ Rikvold,
Electrochim.\ Acta {\bf 36}, 1689 (1991), and work cited therein.

\item{[9]}
S.H.\ Payne, J.~Zhang, and H.J.\ Kreuzer,
Surf.\ Sci.\ {\bf 264}, 185 (1992).

\item{[10]}
R.J.\ Baxter,
{\it Exactly solved models in statistical mechanics}
(Academic, London, 1982).

\item{[11]}
P.A.\ Rikvold, M.~Gamboa-Aldeco, J.~Zhang, M.~Han, Q.~Wang, H.L.\ Richards,
and A.~Wieckowski,
Surf.\ Sci., in press.

\item{[12]}
E.~Domany, M.~Schick, J.S.\ Walker, and R.B.\ Griffiths,
Phys.\ Rev.\ B {\bf 18}, 2209 (1978);
E.~Domany, and M.~Schick, Phys.\ Rev.\ B {\bf 20}, 3828 (1979);
M.~Schick, Prog.\ Surf.\ Sci.\ {\bf 11}, 245 (1981).

\item{[13]}
G.M.\ Buendia, M.A.\ Novotny, and J.~Zhang,
in {\it Computer Simulation Studies in Condensed-Matter Physics VII},
edited by D.P.\ Landau, K.K.\ Mon, and H.-B.\ Sch{\"u}ttler,
Springer Proceedings in Physics {\bf 78}
(Springer, Berlin, 1994).

\item{[14]}
See, {\it e.g.}, K.~Binder, in
{\it Monte Carlo Methods in Statistical Physics, Second Edition},
edited by K.~Binder (Springer, Berlin, 1986).

\item{[15]}
D.P.\ Landau,
Phys.\ Rev.\ B {\bf 27}, 5604 (1983).

\item{[16]}
O.M.\ Magnussen, J.~Hotlos, R.J.\ Nichols, D.M.\ Kolb, and R.J.\ Behm,
Phys.\ Rev.\ Lett.\ {\bf 64}, 2929 (1990).

\item{[17]}
G.J.\ Edens, X.\ Gao, and M.J.\ Weaver,
J.\ Electroanal.\ Chem.\ {\bf 375}, 357 (1994).

\vfill
\eject

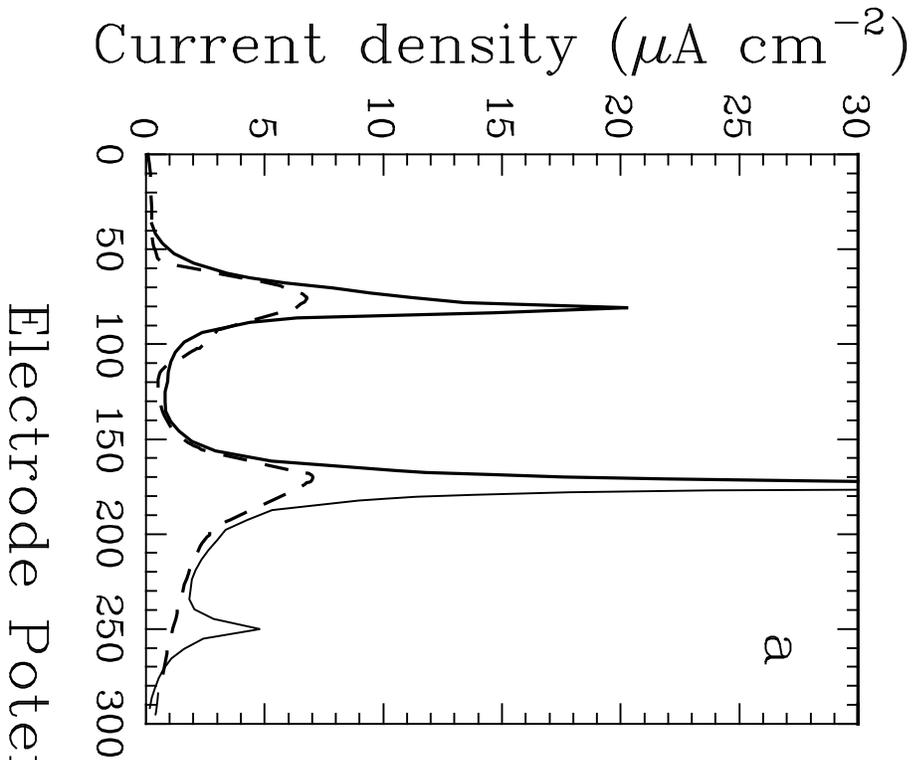
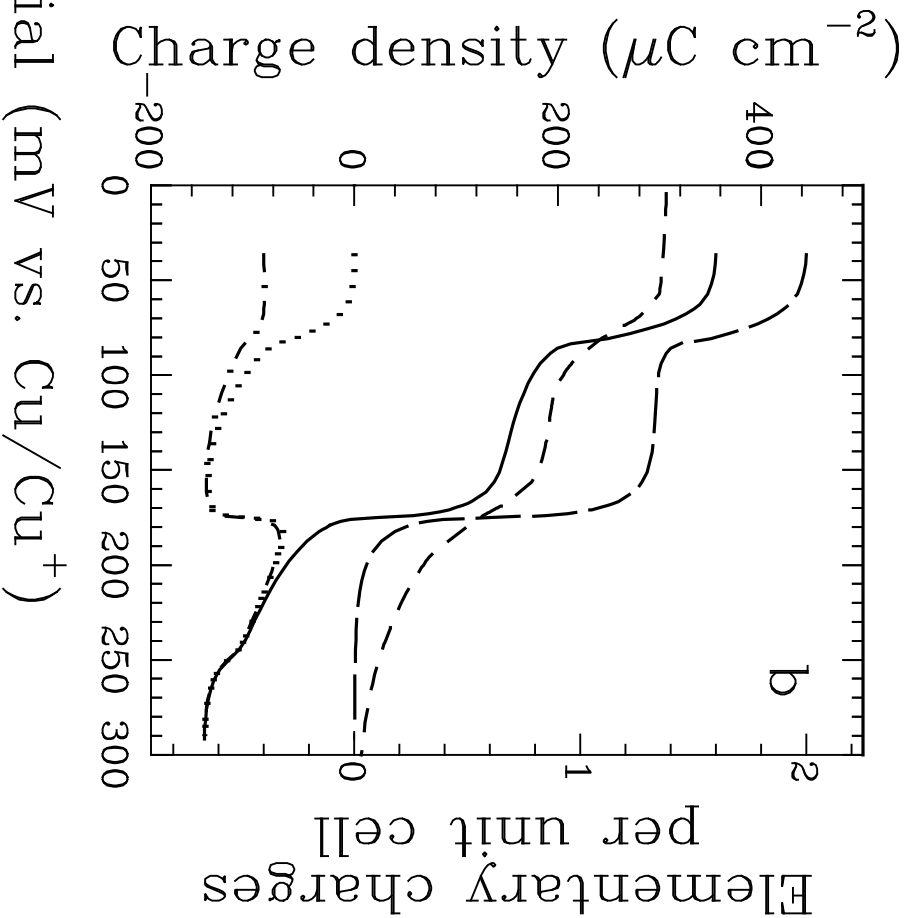
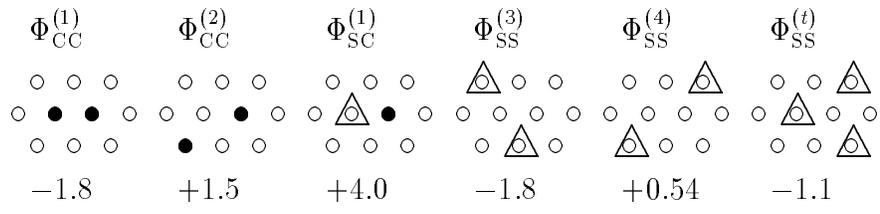
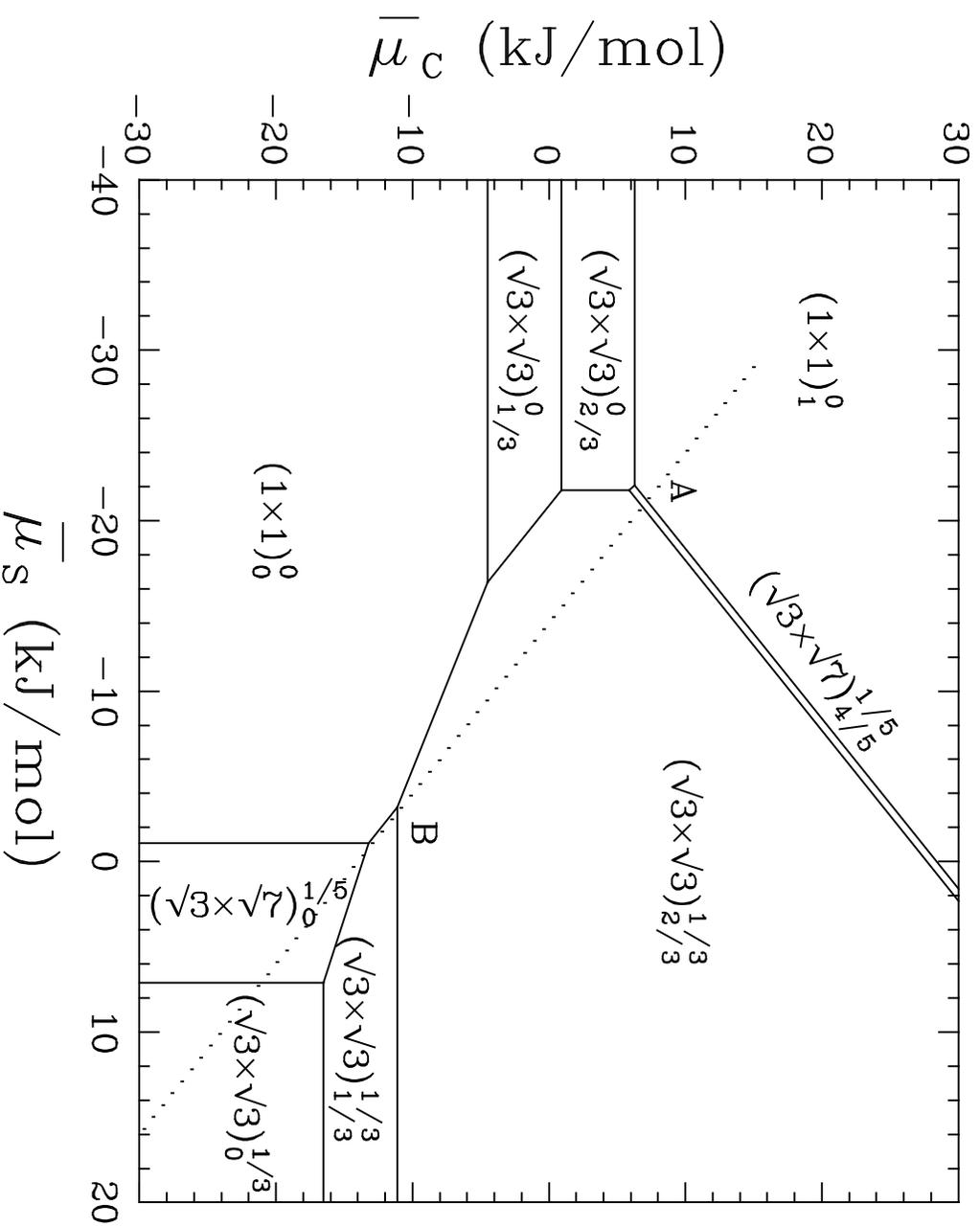
\end